# Graphene – A rising star in view of scientometrics

Andreas Barth* and Werner Marx**

\* FIZ Karlsruhe, D-76344 Eggenstein-Leopoldshafen (Germany)
\** Max Planck Institute for Solid State Research, D-70569 Stuttgart (Germany)

**Abstract**

We have carried out a scientometric analysis of the literature dealing with graphene, a material which has been identified as a new carbon allotrope. The investigation is based on the CAplus database of Chemical Abstracts Service, the INSPEC database of the Institute of Electrical and Electronics Engineers and the Web of Science (WoS) of Thomson Reuters. The time evolution of the publications shows a dramatic increase since 2004 when graphene has been isolated for the first time. The graphene literature has been analyzed with respect to the most productive authors, research organizations, countries of authors, and the leading journals. Furthermore, the time evolution of the major graphene based research topics and the top 10 most productive research organizations with respect to their specific topics have been analyzed and are shown as heat maps. In addition, the graphene research landscape has been established using the analysis tool STN AnaVist®. Finally, a brief citation analysis has been performed. The time evolution of the impact reveals that, rather similar to the output evolution, the impact increase of the graphene papers within the starting period is possibly going to outrun the impact increase of the fullerenes and nanotubes papers.

**Introduction**

Planar one-atom thick single graphite layers of $sp^2$-bonded carbon atoms are named graphene, the newest member of the carbon structural family. Graphene has been called a rising star among new materials [1]. Although it has been discussed since 1947, it was not believed to exist in a free state. The early theoretical studies focused mainly on the analysis of the physical properties of hypothetical two-dimensional crystals. According to Landau and Peierls, two-dimensional crystals are thermodynamically unstable and can not exist in the free state, i.e. without an embedding 3D crystal structure. Until recently, there was no experimental evidence to contradict this hypothesis. In 2004, however, graphene was found unexpectedly by isolation from graphite crystals [2,3] and this discovery required a revision of the theory of 2D crystals and defined a new 2D allotrope of carbon besides diamond and graphite (3D), nanotubes (1D), and fullerenes (0D).

Graphene can be prepared with unique purity and exhibits some remarkable properties: in particular a highly efficient electrical conductivity combined with an extremely fast charge transport and an extraordinary strength. These properties make the material potentially useful in a wide range of applications like in electronics (high speed transistors, one electron transistors) and in materials science (composite materials). In addition, some bizarre predictions of quantum electrodynamics can be verified experimentally much easier than under the extreme conditions typical for astronomy and high energy physics.

The experimental discovery of free-standing graphene sheets as a new member of the carbon structural family caused another "gold rush" around this interesting and promising research field, leading to a strong rise in the number of publications. Since research on graphene has become a "hot topic" for scientists, it is not surprising that the publication (and citation) pattern of such a new research field is also of great interest for scientometric studies. In the study presented on this paper we have analyzed the publication pattern of research on graphene with respect to the specific topics, authors, organizations, and countries. In addition, we had a closer look at the citations.

**Methodology and Information Sources**

The major part of the data presented here is based on the literature database of Chemical Abstracts Service (CAS), a division of the American Chemical Society (ACS). This database is the most extensive source of substance related publications (either articles or patents) in the fields of chemistry, materials science, and also in substance related physics. The database for physics, electronics, and computing (INSPEC) was consulted to distinguish experimental and theoretical work. Access to both databases is provided via the online service STN International [4,5]. The sophisticated STN search system and specific functions for carrying out statistical investigations have made it possible to perform extensive scientometric studies. Further investigations can be performed using STN AnaVist®, an analysis tool developed by STN International. Additional information has been obtained from the Science Citation Index (SCI) which is available both via STN and the Web of Science (WoS). The WoS is the search platform provided by Thomson Reuters [6] (the former Thomson Scientific, emerged from the Institute for Scientific Information (ISI) in Philadelphia). The competent use of such databases and search systems requires some experience and awareness of possibilities and pitfalls.

**Overall productivity: Graphene in Comparison with Nanotubes and Fullerenes**

In the year 2004 Kostya Novoselov and Andre Geim et al., working at the Department of Physics of the University of Manchester (now: Manchester Centre for Mesoscience and Nanotechnology), isolated for the first time a stable two-dimensional crystal of graphite which was later named graphene. The results were published in the prestigious journal *Science* [2] and recently reviewed in *Nature Materials* [1] and *Scientific American* [7]. The bibliographic document of Ref. 2 in CAplus is given in Figure 1.

| | |
|---|---|
| AN | 2004:866953  CAPLUS |
| DN | 142:65915 |
| ED | Entered STN:  20 Oct 2004 |
| TI | Electric Field Effect in Atomically Thin Carbon Films |
| AU | Novoselov, K. S.; Geim, A. K.; Morozov, S. V.; Jiang, D.; Zhang, Y.; Dubonos, S. V.; Grigorieva, I. V.; Firsov, A. A. |
| CS | Department of Physics, University of Manchester, Manchester, M13 9PL, UK |
| SO | Science (Washington, DC, United States) (2004), 306(5696), 666-669<br>CODEN: SCIEAS; ISSN: 0036-8075 |
| PB | American Association for the Advancement of Science |
| DT | Journal |
| LA | English |
| CC | 76-1 (Electric Phenomena) |
| AB | The authors describe monocryst. graphitic films, which are a few atoms thick but are nonetheless stable under ambient conditions, metallic, and of remarkably high quality. The films are a two-dimensional semimetal with a tiny overlap between valence and conductance bands, and they exhibit a strong ambipolar elec. field effect such that electrons and holes in concns. up to 1013 per square centimeter and with room-temperature mobilities of apprx.10,000 square centimeters per V-second can be induced by applying gate voltage. |

**Figure 1:** Bibliographic document of the famous paper by Kostya Novoselov and Andre Geim et al. on the first preparation of stable graphene crystals. Source: CAplus on STN International.

The discussion of the literature time evolution that follows is based on the Science Citation Index (SCI) rather than on the CAS literature database. The number of articles published in the SCI *source journals* has become a standard measure of scientific productivity (output) and allows a comparison with the overall growth of the scientific literature. In addition, the multidisciplinary SCI fits better the interdisciplinary character of the graphene research.

Finally, the WoS is the standard basis for citation data, which are discussed below. Accordingly, the evolution of output and impact can be compared in a more coherent way.

The WoS *General Search* mode reveals publications which appeared in SCI *source journals* (in particular articles, reviews, and meeting abstracts). The publications covered by the SCI do not include books, popular publications, research reports, and conference proceedings (unless they appear in *source journals*). The about 6000 SCI *source journals* carefully selected by the staff of Thomson Reuters as contributing to the progress of science represent only a fraction of the total world wide scientific periodicals, however, these are the dominating journals with respect to availability and impact.

The number of articles of a selected ensemble of publications can easily be plotted versus the publication years using the WoS analyze option. Figure 2 shows the time evolution of the articles of the entire graphene research field. The evolution of the related fullerene and nanotubes research fields is shown for comparison. The total number of articles covered by the SCI is included as a measure for the growth of the overall scientific literature. The term "graphen*" (* = wildcard including the plural) was searched in the WoS title and abstract fields. The fullerene based data were searched in the CAS literature database, using the appropriate CAS Registry numbers. In contrast to graphene and nanotubes, fullerenes are members of a broad substance family with various names and molecular formulas and cannot be selected completely by the search term "fullerene". Note that the term "fullerenes" means here all buckyball related species and not the entire structural family including the carbon nanotubes.

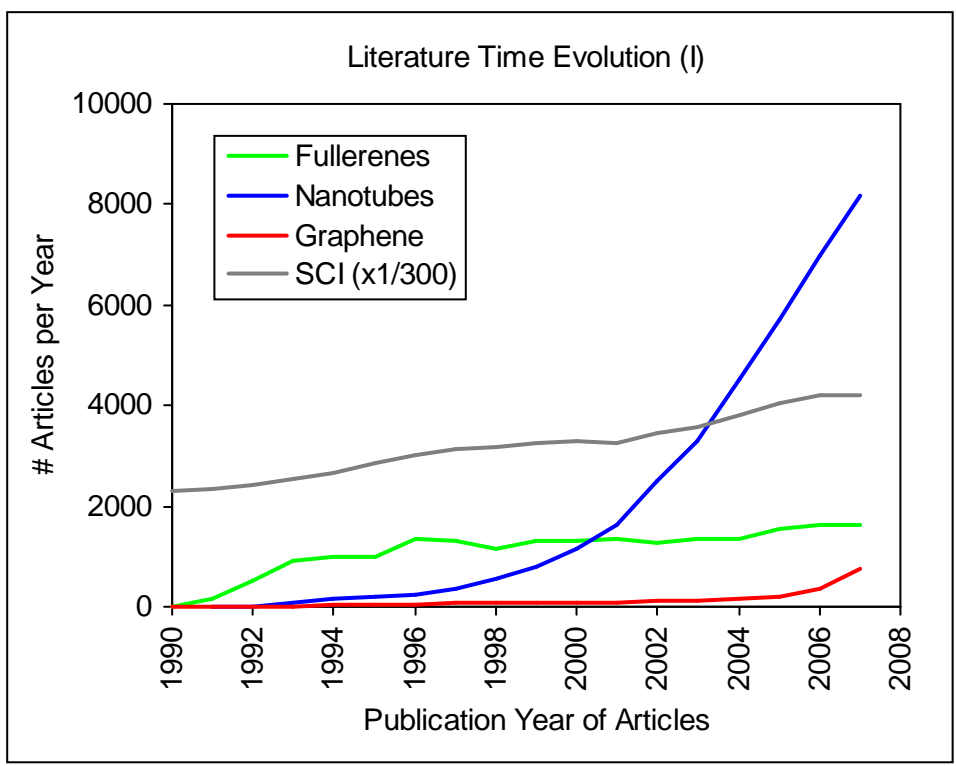

**Figure 2:** Time-dependent number of articles dealing with (1) fullerenes, (2) nanotubes, and (3) graphene. The total number of articles covered by the Science Citation Index (SCI) is shown as a rough measure for the growth of scientific literature. Source: Thomson Reuters Web of Science (WoS) and STN International.

According to Figure 2, the output (i.e. the publication rate expressed as number of articles per year) of the research activities dealing with nanotubes steadily increased, reaching more than 8000 papers published in the year 2007 (compared to "only" 1600 fullerene papers published in the same year). This implies a growth factor of 7 since 2000, which is far above the growth of the overall scientific literature in the same time period (about a factor of 1.25 with respect to the literature appearing in the *source journals* covered by the SCI). In contrast to the nanotubes literature, the time evolution of the fullerene publications shows a distinct saturation since about five years after discovery. The graphene related literature (meanwhile altogether about 3000 articles) increased by almost a factor of 5 since 2004 (compared to a factor of 2 of the nanotubes literature within that same period), obviously starting a follow-up boom beside the ongoing fullerenes and nanotubes reseach activities.

A better comparison of the research activities can be obtained by plotting the time-dependent number of articles within the first decade after the discovery. For the sake of convenience, the beginning of the fullerene research is assumed to be the discovery of fullerene material by Wolfgang Krätschmer published 1990 [8] and the beginning of the nanotubes research is assumed to be the discovery of nanotubes by Sumio Ijima published 1991 [9]. Figure 2 shows the literature evolution within the corresponding time periods of one decade.

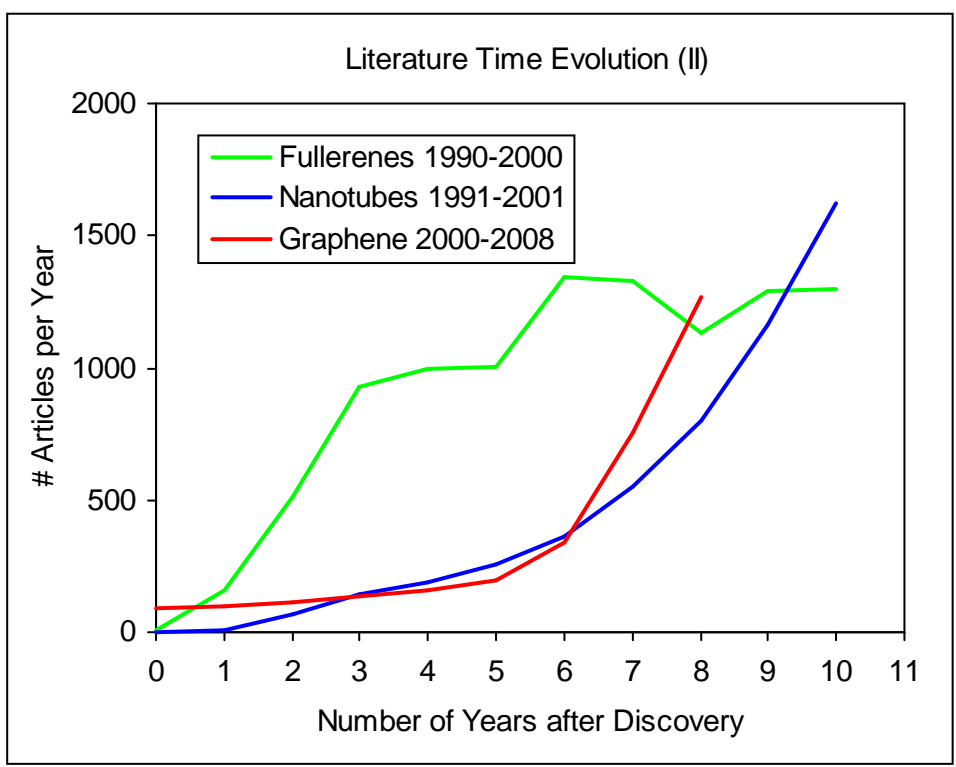

**Figure 3:** Time-dependent number of articles within the first decade after discovery dealing with (1) fullerenes, (2) nanotubes, and (3) graphene. The number of the graphene articles in the year 2008 has been extrapolated from the available data at the end of May. Source: Thomson Reuters Web of Science (WoS).

According to Figure 3, research around fullerenes shows a strong increase in the first 3 years while both nanotubes and graphene show a delayed take-off. Between 1990 and 2003 the literature mentioning the term graphene reached 778 articles, followed by a strong increase after the discovery of Novoselov and Geim in the year 2004, amounting to a total number of currently 2683 articles. Extrapolating the number of articles published till end of May results in almost 1300 papers for the publication year 2008. The extrapolation is based on the assumption that the WoS input time delay is about one month, i.e. that at the end of May of the current year about 1/3 of the 2008 publications are included.

The documents covered by the INSPEC database enable the selection of papers with a focus on experimental and theoretical research, respectively. In Figure 4 the time evolution of the publications of these two categories is shown. The total number of publications from 1968 till 2007 with the terms "graphen*" in the title or abstract amounts to 1731 papers (experimental: 1070, theoretical: 771). Although graphene has been initially more the research subject of theoreticians, it is interesting to note that in most years the number of experimental studies has outnumbered the theoretical studies. Since 2006, however, the situation reversed. Note that some publications have been assigned as both experimental and theoretical resulting in a small overlap.

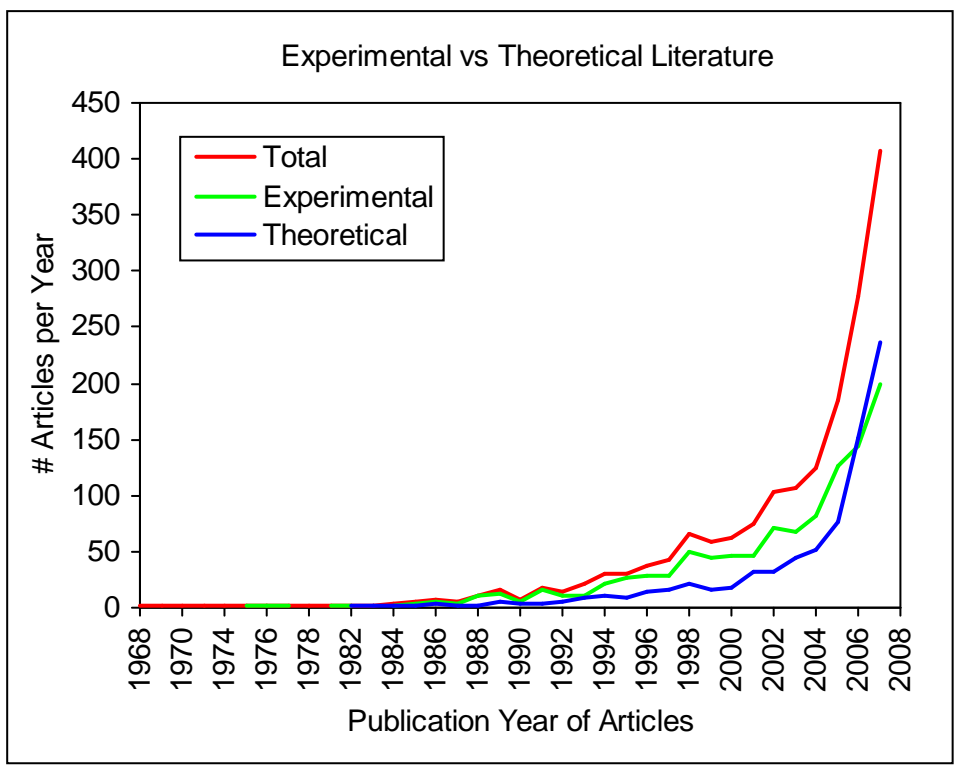

**Figure 4:** Time-dependent number of articles with respect to experimental and theoretical work. Note: Some publications have been assigned as both experimental and theoretical resulting in a small overlap. Source: INSPEC on STN International.

**Productivity: Authors, Research Organizations, Journals**

Compared to the SCI (and INSPEC), the CAS literature database offers some advantages for an in-depth analysis with regard to the specific research topics within graphene research. Therefore, the following data were established using the CAS rather than the SCI database. At the date of search (11-04-2008) the total number of publications covered by the CAS literature database CAplus mentioning graphene in the title or the abstract amounts to roughly 3.500 research articles (including preprint server documents not covered by the WoS, about 100 patents were excluded). The answer set has been analyzed with respect to the most productive authors, research organizations, countries, and the leading journals. The results of this analysis are shown in Tables 1-4. Note that in the CAS literature database the countries of authors and their affiliations are assigned only to the first authors of the papers (and not to all authors as in the WoS records).

**Table 1:** Top 10 most productive authors with respect to the number of articles dealing with graphene. Source: CAplus on STN International.

| # | Author | Country | # Articles | % Articles |
|---|---|---|---|---|
| 1 | Guinea F | Spain | 86 | 2.5 |
| 2 | Castro Neto A H | USA | 71 | 2.1 |
| 3 | Katsnelson M I | Netherlands | 57 | 1.7 |
| 4 | Peres N M R | Portugal | 57 | 1.7 |
| 5 | Geim A K | UK | 52 | 1.5 |
| 6 | Endo M | Japan | 43 | 1.2 |
| 7 | Das Sarma S | USA | 42 | 1.2 |
| 8 | Novoselov K S | UK | 41 | 1.2 |
| 9 | Beenakker C W J | Netherlands | 36 | 1.0 |
| 10 | Dresselhaus M S | USA | 35 | 1.0 |

The most productive author is Francisco Guinea (USA) with 86 papers dealing with graphene and comprising 2.5% of all papers published in this research field. The authors of the seminal publication on graphene given in Figure 1 [2], Geim and Novoselov, appear on rank 5 (52 papers) and 8 (41 papers), respectively.

**Table 2:** Top 10 most productive research organizations with respect to the number of articles dealing with graphene. Source: CAplus on STN International.

| # | Research Org. | # Articles | % Articles |
|---|---|---|---|
| 1 | Univ. of California | 87 | 2.5 |
| 2 | Penn. State Univ | 52 | 1.5 |
| 3 | Univ. of Maryland | 44 | 1.3 |
| 4 | Tokyo Inst. of Techn. | 43 | 1.3 |
| 5 | Chin. Acad. Sci. | 43 | 1.3 |
| 6 | Consejo Sup. Invest. Cient. | 42 | 1.2 |
| 7 | MIT | 41 | 1.2 |
| 8 | Columbia Univ. | 40 | 1.2 |
| 9 | Boston Univ. | 40 | 1.2 |
| 10 | MPI Metals Research | 36 | 1.1 |

Among the top 10 research organizations there are 6 from USA, 1 from Japan, Peop. Rep. China, Spain, and Germany each. It is interesting to note that the top author (Francisco Guinea, see Table 1) has published almost as many papers as the top research organization (Univ. of California) altogether.

**Table 3:** Top 10 most productive countries of authors with respect to the number of articles dealing with graphene. Source: CAplus on STN International.

| # | Author Country | # Articles | % Articles |
|---|---|---|---|
| 1 | USA | 1106 | 30.6 |
| 2 | Japan | 514 | 14.2 |
| 3 | Germany | 215 | 6.0 |
| 4 | Peop. Rep. China | 199 | 5.5 |
| 5 | France | 181 | 5.0 |
| 6 | UK | 144 | 4.0 |
| 7 | Spain | 137 | 3.8 |
| 8 | Netherlands | 99 | 2.7 |
| 9 | Russia | 96 | 2.7 |
| 10 | Italy | 90 | 2.5 |

The top 10 countries comprise 77% of all publications worldwide and the top 3 countries published already about 50% of all publications. It is interesting to compare the number of publications of the top 10 countries with the number of publications of the top 10 authors. The first author in our ranking has almost as many publications as country number 10 (Italy). The number of publications of the top 10 authors altogether is equal to the number of publications of the second country in our ranking (Japan).

**Table 4:** Leading journals with respect to the number of articles dealing with graphene. Source: CAplus on STN International.

| # | Source Journal | # Articles | % Articles |
|---|---|---|---|
| 1 | LOS ALAMOS NATIONAL LABORATORY, PREPRINT ARCHIVE, CONDENSED MATTER | 707 | 19.4 |
| 2 | PHYS. REV. B: CONDENSED MATTER MATER. PHYS. | 416 | 11.4 |
| 3 | CARBON | 182 | 5.0 |
| 4 | PHYS. REV. LETT. | 173 | 4.7 |
| 5 | ARXIV.ORG, E-PRINT ARCHIVE, CONDENSED MATTER | 134 | 3.7 |
| 6 | APPL. PHYS. LETT. | 92 | 2.5 |
| 7 | NANO LETT. | 50 | 1.4 |
| 8 | CHEM. PHYS. LETT. | 46 | 1.3 |
| 9 | AIP CONF. PROC. | 40 | 1.1 |
| 10 | J. PHYS. CHEM. B | 36 | 1.0 |
| 11 | NANOTECHNOLOGY | 28 | 0.8 |
| 12 | LANGMUIR | 20 | 0.6 |

The "journal" ranking includes two preprint servers (1,5) and one conference proceedings (9), all other entries are regular research journals. Due to current publishing habits, there is some overlap of publications from preprint servers and journals, effecting somewhat the publication counting in the Tables 1-4 and in the heat maps below.

**Heat Maps**

A further analysis of the publications with respect to two parameters can be performed using a two-field co-occurrence chart. In our study we have analyzed the time evolution of the major research topics (see Table 5) and the top 10 most productive research organizations with respect to their specific topics (see Table 6), both within the graphene research field.

**Table 5:** Time-dependent number of publications since 2003 with respect to the top 20 subjects within the graphene research field. The presentation of the data as a heat map allows the identification of hot topics.

|  | 2008 | 2007 | 2006 | 2005 | 2004 | 2003 | SUM | TOTAL | PERC. |
|---|---|---|---|---|---|---|---|---|---|
| Nanotubes | 15 | 124 | 106 | 103 | 77 | 69 | 494 | 973 | 51% |
| Electric conductivity | 12 | 188 | 75 | 18 | 6 | 2 | 301 | 590 | 51% |
| Band structure | 16 | 123 | 52 | 20 | 11 | 10 | 232 | 448 | 52% |
| Density of states | 17 | 122 | 57 | 19 | 11 | 11 | 237 | 457 | 52% |
| Nanostructures | 20 | 119 | 40 | 27 | 21 | 12 | 239 | 458 | 52% |
| Simulation and Modeling | 11 | 77 | 35 | 13 | 16 | 8 | 160 | 309 | 52% |
| Quantum Hall effect | 1 | 96 | 61 | 12 | 1 | 0 | 171 | 341 | 50% |
| Electronic structure | 6 | 76 | 30 | 10 | 11 | 1 | 134 | 262 | 51% |
| Magnetic field effects | 2 | 95 | 45 | 3 | 1 | 0 | 146 | 290 | 50% |
| Carbon fibers | 3 | 12 | 11 | 17 | 20 | 11 | 74 | 145 | 51% |
| Band gap | 11 | 76 | 20 | 8 | 4 | 3 | 122 | 233 | 52% |
| Surface structure | 5 | 43 | 19 | 14 | 8 | 13 | 102 | 199 | 51% |
| Density functional theory | 4 | 53 | 17 | 6 | 9 | 13 | 102 | 200 | 51% |
| Fermions | 0 | 77 | 33 | 5 | 0 | 0 | 115 | 230 | 50% |
| Raman spectra | 5 | 39 | 16 | 8 | 10 | 6 | 84 | 163 | 52% |
| Vapor deposition process | 4 | 15 | 15 | 17 | 18 | 12 | 81 | 158 | 51% |
| Electric current carriers | 6 | 66 | 20 | 5 | 0 | 2 | 99 | 192 | 52% |
| Microstructure | 3 | 23 | 9 | 7 | 10 | 13 | 65 | 127 | 51% |
| Fermi level | 5 | 46 | 20 | 7 | 2 | 4 | 84 | 163 | 52% |
| Disorder | 3 | 38 | 30 | 7 | 2 | 1 | 81 | 159 | 51% |
| Intercalation | 1 | 4 | 7 | 6 | 6 | 9 | 33 | 65 | 51% |
| Quasiparticles & Excitations | 2 | 46 | 29 | 5 | 2 | 1 | 85 | 168 | 51% |
| Electric resistance | 4 | 48 | 11 | 3 | 2 | 2 | 70 | 136 | 51% |
| Adsorption | 0 | 19 | 23 | 10 | 4 | 7 | 63 | 126 | 50% |
| Electron density | 4 | 41 | 14 | 2 | 4 | 4 | 69 | 134 | 51% |
| Battery anodes | 1 | 2 | 5 | 3 | 6 | 9 | 26 | 51 | 51% |
| Landau level | 0 | 57 | 21 | 0 | 0 | 0 | 78 | 156 | 50% |
| Fullerenes | 1 | 6 | 11 | 7 | 2 | 2 | 29 | 57 | 51% |
| Magnetoresistance | 1 | 38 | 21 | 4 | 2 | 0 | 66 | 131 | 50% |
| Nanoparticles | 0 | 21 | 13 | 10 | 8 | 4 | 56 | 112 | 50% |
| SUM | 163 | 1790 | 866 | 376 | 274 | 229 | 3698 | 7233 |  |
| TOTAL | 545 | 5258 | 2550 | 1243 | 936 | 802 |  |  |  |
| PERC. | 30% | 34% | 34% | 30% | 29% | 29% |  |  |  |

From this table we get a better inside into the pattern of the research on graphene. Colored cells show the number of publications of a specific research field in a given year (red: high number, blue: low number). Additional columns show the sum of publications of the years presented in this table (SUM) and the total sum over all years (TOTAL). The last column shows the percentages of the number of publications in the time period 2003-2008 with respect to all publications. Similarly, the same is valid for rows. However, in the case of research fields (rows) a publication may show up in more than one research area.

According to the percentages in the last column, more than 50% of all publications on graphene with respect to the major research topics have been published in the last 5 years. More than 30% of all publications in a given year have a focus on one (or more) of the top 20 research subjects. Looking at the trends one can see that several research areas show a strong increase in the number of publications. Among the major research topics showing more than a doubling of the publications between 2006 and 2007 are: electric conductivity, band structure, density of states, nanostructures, simulation and modeling, electronic structure, magnetic field effects, band gap, surface structure, density functional theory, fermions, raman spectra, electric current carriers, microstructure, Fermi level, electric resistance, electron density, and Landau level.

**Table 6:** Number of publications of the top 10 most productive research organizations with respect to the top 20 subjects within the graphene research field. The presentation of the data as a heat map allows the identification of hot topics.

| | Univ. of California | Penn. State Univ. | Univ. Sys. of Maryland | Tokyo Inst. of Techn. | Chin. Acad. of Sci. | Consejo Sup. de Invest. Cient. | MIT | Columbia Univ. | MPI for Metal Res. | Boston Univ. | SUM | TOTAL | PERC. |
|---|---|---|---|---|---|---|---|---|---|---|---|---|---|
| Nanotubes | 7 | 8 | 2 | 6 | 14 | 2 | 6 | 2 | 5 | 0 | 52 | 703 | 7% |
| Electric conductivity | 5 | 2 | 11 | 5 | 1 | 6 | 4 | 9 | 4 | 5 | 52 | 316 | 16% |
| Band structure | 7 | 1 | 2 | 6 | 3 | 3 | 4 | 5 | 1 | 2 | 34 | 265 | 13% |
| Density of states | 2 | 4 | 1 | 8 | 1 | 12 | 2 | 2 | 6 | 10 | 48 | 264 | 18% |
| Nanostructures | 10 | 3 | 1 | 3 | 2 | 1 | 0 | 4 | 1 | 1 | 26 | 249 | 10% |
| Simulation and Modeling | 5 | 2 | 4 | 3 | 1 | 1 | 1 | 1 | 0 | 5 | 23 | 189 | 12% |
| Quantum Hall effect | 11 | 1 | 1 | 2 | 2 | 1 | 9 | 13 | 0 | 7 | 47 | 170 | 28% |
| Electronic structure | 3 | 1 | 0 | 4 | 0 | 7 | 1 | 0 | 1 | 5 | 22 | 160 | 14% |
| Magnetic field effects | 0 | 3 | 0 | 2 | 0 | 5 | 3 | 10 | 1 | 2 | 26 | 149 | 17% |
| Band gap | 7 | 3 | 0 | 3 | 1 | 0 | 0 | 3 | 2 | 4 | 23 | 135 | 17% |
| Carbon fibers | 1 | 1 | 0 | 2 | 1 | 0 | 3 | 0 | 1 | 0 | 9 | 132 | 7% |
| Surface structure | 5 | 0 | 1 | 1 | 0 | 4 | 0 | 2 | 3 | 0 | 16 | 126 | 13% |
| Density functional theory | 3 | 2 | 0 | 0 | 0 | 1 | 3 | 0 | 0 | 1 | 10 | 119 | 8% |
| Fermions | 10 | 5 | 1 | 0 | 0 | 4 | 2 | 2 | 1 | 7 | 32 | 115 | 28% |
| Raman spectra | 5 | 3 | 0 | 4 | 4 | 0 | 2 | 6 | 1 | 0 | 25 | 114 | 22% |
| Vapor deposition proc. | 0 | 1 | 0 | 0 | 1 | 0 | 0 | 2 | 0 | 0 | 4 | 107 | 4% |
| Electric current carriers | 4 | 0 | 11 | 3 | 0 | 0 | 0 | 6 | 2 | 4 | 30 | 105 | 29% |
| Microstructure | 3 | 1 | 0 | 1 | 1 | 0 | 0 | 0 | 4 | 0 | 10 | 97 | 10% |
| Fermi level | 1 | 2 | 1 | 4 | 1 | 3 | 1 | 2 | 1 | 1 | 17 | 95 | 18% |
| Disorder | 6 | 2 | 1 | 5 | 0 | 4 | 1 | 1 | 2 | 13 | 35 | 93 | 38% |
| SUM | 95 | 45 | 37 | 62 | 33 | 54 | 42 | 70 | 36 | 67 | | | |
| TOTAL | 394 | 229 | 186 | 194 | 154 | 200 | 158 | 194 | 161 | 214 | | | |
| PERC. | 24% | 20% | 20% | 32% | 21% | 27% | 27% | 36% | 22% | 31% | | | |

Table 6 shows the top 10 research organizations with respect to the top 20 research subjects. The colored cells contain the number of publications (red: high number, blue: low number). Due to cooperation between the research organizations, some publications may occur in more than one column. Since research may focus on several research fields, some publications may occur in more than one row.

The top 10 research organizations publish with a strong focus in the top 20 research subjects, although the research may include an emphasis on other research areas not listed in this table. Within the set of the top 20 research subjects, some research organizations show a stronger focus (with more than 10 publications in at least 2 of the research fields): University of California, University of Maryland, Columbia University, and Boston University.

**Research Landscape**

The online service STN International has recently launched a new interactive analysis tool called STN AnaVist® [10] which focuses mainly on patent analysis [11]. This tool can also be applied to literature exploring the usefulness for basic research fields. One of the functions offered by STN AnaVist® implies the creation of so-called research landscapes: "Significant keywords and concepts are derived from document titles and abstracts. These keywords are used to determine the similarity between documents. An algorithm uses document similarity scores to position each document relative to one another in a two-dimensional space, with each document positioned at one point. This process is repeated until all documents have been clustered and each assigned to a single x, y coordinate pair. A graphical map is generated. The z coordinates, determining the height of each 'peak', are calculated based on the density of the documents in an area."

In Figure 5 the research landscape of the graphene publications covered by the CAS literature file is shown. The green areas correspond to different clusters of documents. The clusters are characterized by the two top clustering words, e. g. "grow, catalyst" at the top of the landscape or "quanta, landau" at the bottom. Areas with a violet top comprise clusters with a rather large number of publications. According to the clustering concept, major areas of research include "grow, catalyst", "nanotube, carbon nanotube", "surface, adsorption", "edge, ribbon", and "quanta, landau".

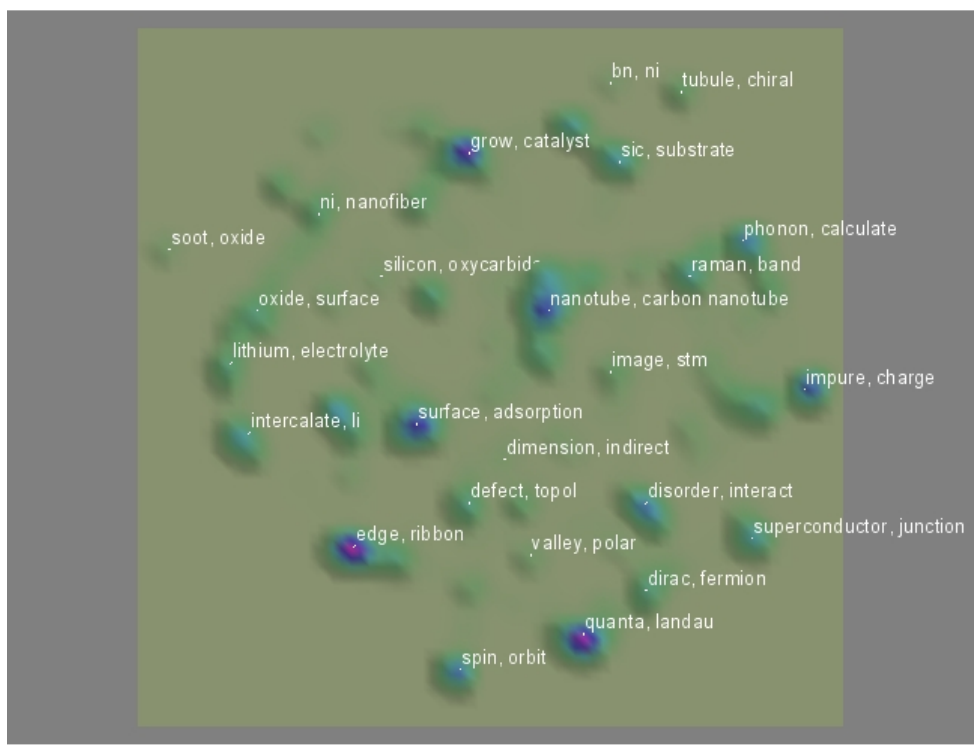

**Figure 5:** Research landscape of graphene publications established by the analysis tool STN AnaVist®. Source: CAplus on STN International.

In an interactive mode it is possible to zoom into the research landscape and perform a more detailed analysis. As an example, we show in Figure 6 a zoom of the clusters around "quanta, landau" in Figure 5 with the quantum hall effect as the major research subject. This view shows a more detailed picture with various new clusters not visible on the full map.

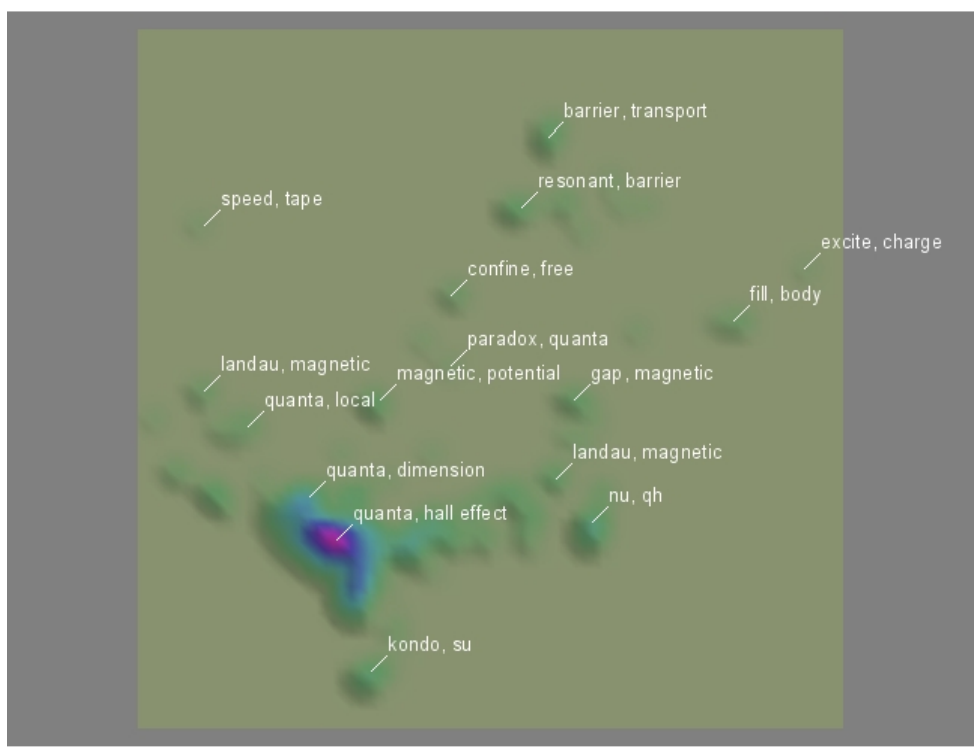

**Figure 6:** Research landscape of graphene publications focusing on the clusters around the quantum hall effect established by the analysis tool STN AnaVist®. Source: CAplus on STN International.

**Citation Analysis**

The number of citations is often taken as a measure of the attention an article, a researcher, an institute or even a country has attracted. Although citation numbers are no ultimate scale of the final importance and quality of articles, they reflect strengths and shortcomings and are therefore frequently used for research evaluation. Being cited means that a given article appears as a reference in the article of another author for additional reading. The number of citations is thus a rough measure of the importance or usefulness of the paper within the scientific community.

Although the CAS literature database meanwhile offers the functionality of a citation index, the SCI based citation data are still the standard for measuring impact. The impact data corresponding to the output data of the three related research fields given in Figure 3 have been established under the WoS and are shown in Figure 7. The time evolution of the overall number of citations reveals that the impact increase of the graphene papers is possibly going to outrun the increase of the other two fields, rather similar to the output evolution. Extrapolating the number of citations of the articles published till May of the current year results in about 25,000 citations of the graphene articles for the publication year 2008. Again it is assumed that the WoS input time delay is about one month, i.e. that at the end of May about 1/3 of the citations within the year 2008 are included.

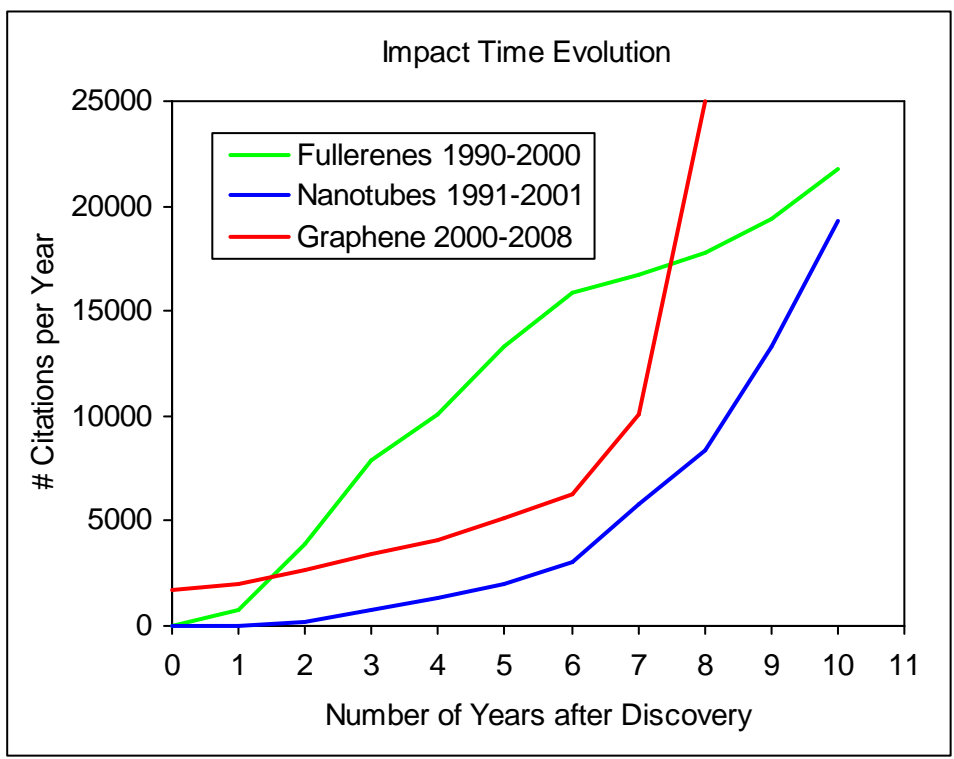

**Figure 7:** Time-dependent number of citations within the first decade after discovery of the articles dealing with (1) nanotubes, (2) fullerenes, and (3) graphene. The number of citations of the graphene articles within the year 2008 has been extrapolated from the available data at the end of May. Source: Thomson Reuters Web of Science (WoS).

Using the recently developed HistCite[®] software of Eugene Garfield, we established the citation graph shown in Figure 8, which is based on the 2514 graphene articles accessible in WoS at the date of search (11-04-2008). This graph visualizes the citation network within the graphene literature published so far: The nodes represent the graphene papers with at least 50 citations collected within the ensemble of the selected graphene articles (local citation score, LCS). Lower citation limits increase the number of nodes considerably without changing the node pattern substantially. The specific papers represented by the circles of the citation graph are given in the table below the graph in short form. Like AnaVist[®], also HistCite[®] is used in an interactive mode and can deliver much more information than given in the limited citation graph of Figure 8.

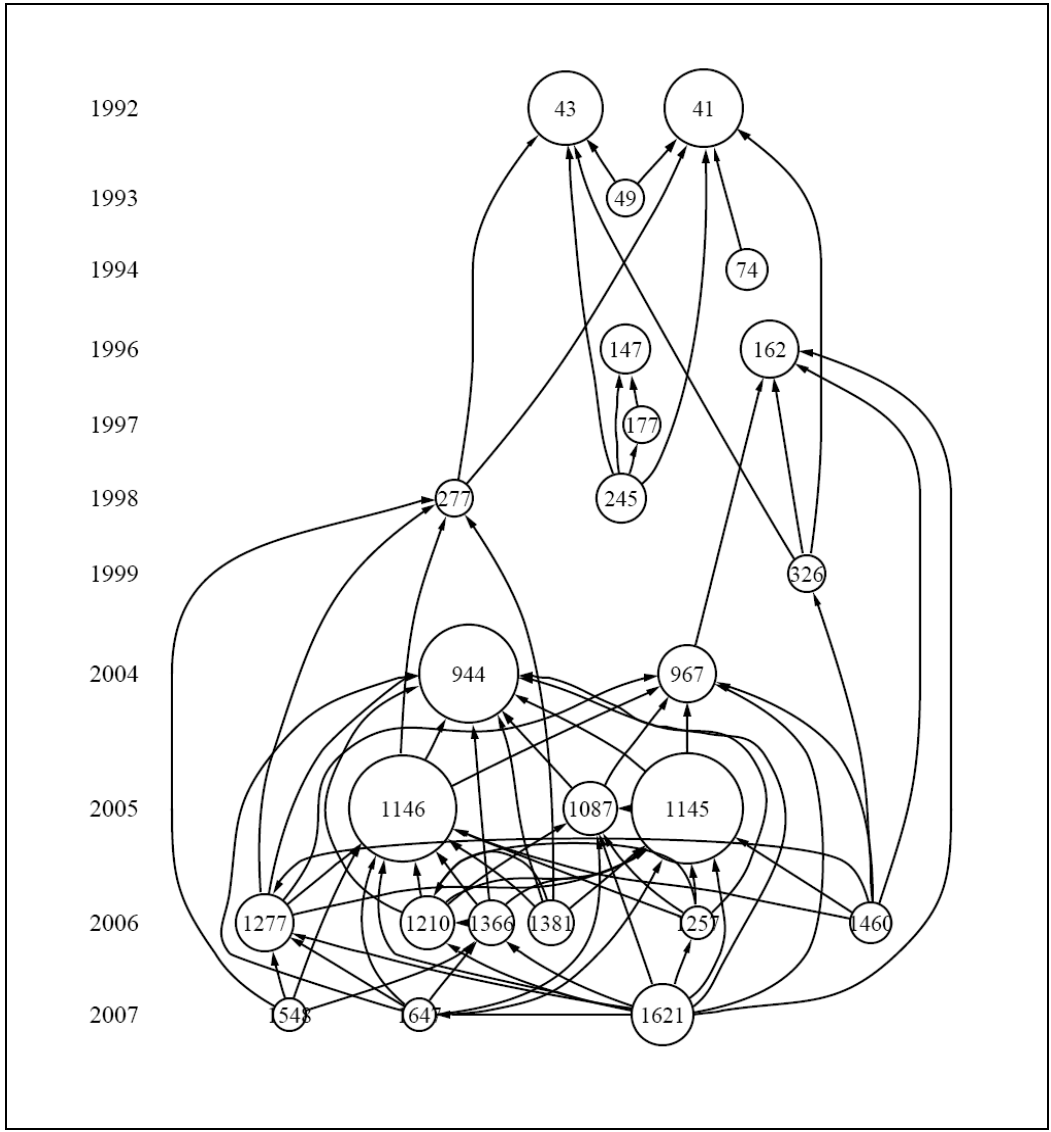

**Figure 8:** Citation graph based on the currently 2514 graphene articles (WoS) established by using the HistCite® software of Eugene Garfield. The nodes represent the graphene papers with at least 50 citations collected within the ensemble of the selected graphene articles (local citation score, LCS). The circle diameter is proportional to the number of citations and the arrows indicate the citation direction between the papers. The numbers within the circles originate from the consecutive numbering of the 2514 graphene papers by the software and do not represent their citation counts.

Nodes: 24, Links: 72
LCS >= 50; Min: 53, Max: 583 (LCS scaled)

|    |     |                                          | LCS | GCS  |
|----|-----|------------------------------------------|-----|------|
| 1. | 41  | Saito R, 1992, APPL PHYS LETT, V60, P2204 | 309 | 1112 |
| 2. | 43  | Saito R, 1992, PHYS REV B, V46, P1804     | 263 | 630  |
| 3. | 49  | Saito R, 1993, J APPL PHYS, V73, P494     | 69  | 161  |
| 4. | 74  | Blasé X, 1994, PHYS REV LETT, V72, P1878  | 85  | 458  |
| 5. | 147 | Thess A, 1996, SCIENCE, V273, P483        | 127 | 2366 |
| 6. | 162 | Nakada K, 1996, PHYS REV B, V54, P17954   | 160 | 296  |
| 7. | 177 | Rao AM, 1997, SCIENCE, V275, P187         | 65  | 1044 |
| 8. | 245 | Odom TW, 1998, NATURE, V391, P62          | 126 | 976  |

| | | | | LCS | GCS |
|---|---|---|---|---|---|
| 9. | 277 | Ando T, 1998, J PHYS SOC JPN, V67, P2857 | | 68 | 139 |
| 10. | 326 | Wakabayashi K, 1999, PHYS REV B, V59, P8271 | | 75 | 154 |
| 11. | 944 | Novoselov KS, 2004, SCIENCE, V306, P666 | | 480 | 536 |
| 12. | 967 | Berger C, 2004, J PHYS CHEM B, V108, P19912 | | 152 | 160 |
| 13. | 1087 | Novoselov KS, 2005, PROC NAT ACAD SCI USA, V102, P10451 | | 140 | 154 |
| 14. | 1145 | Novoselov KS, 2005, NATURE, V438, P197 | | 583 | 611 |
| 15. | 1146 | Zhang YB, 2005, NATURE, V438, P201 | | 557 | 576 |
| 16. | 1210 | Novoselov KS, 2006, NAT PHYS, V2, P177 | | 140 | 142 |
| 17. | 1257 | Katsnelson MI, 2006, EUR PHYS J B, V51, P157 | | 60 | 65 |
| 18. | 1277 | Berger C, 2006, SCIENCE, V312, P1191 | | 171 | 178 |
| 19. | 1366 | Ohta T, 2006, SCIENCE, V313, P951 | | 102 | 107 |
| 20. | 1381 | Katsnelson MI, 2006, NAT PHYS, V2, P620 | | 101 | 106 |
| 21. | 1460 | Son YW, 2006, NATURE, V444, P347 | | 82 | 92 |
| 22. | 1548 | Bostwick A, 2007, NAT PHYS, V3, P36 | | 53 | 55 |
| 23. | 1621 | Geim AK, 2007, NAT MATER, V6, P183 | | 191 | 201 |
| 24. | 1647 | Meyer JC, 2007, NATURE, V446, P60 | | 59 | 63 |

The citation graph of Figure 8 reveals two unequally pronounced clusters of graphene related publications: the pre-2004 articles being less cross-linked and the past-2004 papers being strongly networked with some of them being heavily cited already a few years after their publication. The papers with at least 50 citations within the ensemble of the graphene publications (local citation score, LCS) and their overall number of citations (global citation score, GCS) are given in the table below the graph. One can speculate that the large difference between LCS and GCS of the pre-2004 papers is caused by their broader focus (in particular with respect to nanotubes). The currently appearing articles dealing with graphene discuss the new material as the main topic.

**Summary**

In this study the literature on graphene, a promising new material, has been analyzed by scientometric methods. The time evolution of the literature shows a strong increase since the publication of the first isolation of graphene sheets from graphite crystals in 2004. Similarly, the time evolution of the overall number of citations reveals that the impact increase of the graphene papers is possibly going to outrun the impact increase of the related research fields on fullerenes or nanotubes. The most productive author (Francisco Guinea, USA) has published 86 papers comprising 2.5% of all papers published in the graphene research field. Among the top 10 most productive research organizations there are 6 from USA, 1 from Japan, Peop. Rep. China, Spain, and Germany each. The top 10 most productive countries comprise 77% of all publications worldwide and the top 3 countries alone have published already about 50% of all publications. Time-dependent research patterns are obtained from a heat map showing that about 50% of all articles on graphene with respect to the major research topics have been published in the last 5 years. The top 10 most productive research organizations publish with a strong focus in the top 20 research subjects. A research landscape has been established illustrating the major research clusters with regard to the clustering concept. Finally, a citation graph has been constructed revealing two unequally pronounced clusters of graphene related publications: the pre-2004 articles being less cross-linked and the past-2004 papers being strongly networked with some of them being heavily cited already a few years after their publication.

**References and Information Sources**

[1] A.K. Geim and K.S. Novoselov, The rise of graphene, Nature Materials 6, 183-191 (2007).

[2] K.S. Novosolov et al., Electric field effect in atomically thin carbon films,
Science 306, 666-669 (2004).

[3] K.S. Novosolov et al., Two-dimensional atomic crystals,
Proc. Natl. Acad. Sci. USA 102, 10451-10453 (2005).

[4] http://www.stn-international.de/stndatabases/databases/caplus.html

[5] http://www.stn-international.de/stndatabases/databases/inspec.html

[6] URL Thomsonr Reuters: http://scientific.thomsonreuters.com/products/wos/

[7] A.K. Geim and P. Kim, Carbon wonderland, Scientific American 298, 90-97 (2008).

[8] W. Krätschmer, L.D. Lamb, K, Fostiropoulos, and D.R. Huffman,
Solid $C_{60}$ – a new form of carbon, Nature 347, 354-358 (1990).

[9] S. Iijima, Helical microtubules of graphitic carbon, Nature 354, 56-58 (1991).

[10] http://www.stn-international.de/stninterfaces/stnanavist/stn_anavist.html

[11] G. Fischer and N. Lalyre, Analysis and visualisation with host-based software –
The features of STN® AnaVist™, World Patent Information 28, 312-318 (2006).

Corresponding author: Dr. Werner Marx
Max Planck Institute for Solid State Research
Heisenbergstraße 1, D-70569 Stuttgart, Germany
E-mail: w.marx@fkf.mpg.de

E-mail Dr. Andreas Barth: andreas.barth@fiz-karlsruhe.de